\documentclass[10pt]{article}
\usepackage[utf8]{inputenc}
\usepackage[T1]{fontenc}
\usepackage{lineno}
\usepackage[
 left=2.5cm,
 right=2.5cm,
 top=2.5cm,
bottom=2.5cm,]{geometry}
\usepackage{amsmath}
\usepackage{graphicx}
\usepackage{authblk}
\usepackage[square,numbers]{natbib}
\usepackage{xurl}
\title{Deep Learning for Motion Classification in Ankle Exoskeletons Using Surface EMG and IMU Signals}

\author[1,2]{Silas {Ruhrberg Estévez}}
\author[1]{Josée Mallah}
\author[1]{Dominika Kazieczko}
\author[1]{Chenyu Tang}
\author[1,*]{Luigi G. Occhipinti}

\affil[1]{Electrical Engineering Division, Department of Engineering, University of Cambridge, Cambridge CB3 0FA,
UK}
\affil[2]{School of Clinical Medicine, University of Cambridge, Cambridge CB2 0SP,  UK}

\affil[*]{lgo23@cam.ac.uk}

\date{}

\begin{document}

\maketitle
\begin{abstract}
Ankle exoskeletons have garnered considerable interest for their potential to enhance mobility and reduce fall risks, particularly among the aging population. The efficacy of these devices relies on accurate real-time prediction of the user's intended movements through sensor-based inputs. This paper presents a novel motion prediction framework that integrates three Inertial Measurement Units (IMUs) and eight surface Electromyography (sEMG) sensors to capture both kinematic and muscular activity data. A comprehensive set of activities, representative of everyday movements in barrier-free environments, was recorded for the purpose. Our findings reveal that Convolutional Neural Networks (CNNs) slightly outperform Long Short-Term Memory (LSTM) networks on a dataset of five motion tasks, achieving classification accuracies of $96.5 \pm 0.8 \%$ and $87.5 \pm 2.9 \%$, respectively. Furthermore, we demonstrate the system's proficiency in transfer learning, enabling accurate motion classification for new subjects using just ten samples per class for finetuning. The robustness of the model is demonstrated by its resilience to sensor failures resulting in absent signals, maintaining reliable performance in real-world scenarios. These results underscore the potential of deep learning algorithms to enhance the functionality and safety of ankle exoskeletons, ultimately improving their usability in daily life.
\end{abstract}

\section*{Introduction}

Globally, approximately one in ten people is over the age of $65$, a proportion that is expected to double by the end of the century \cite{Gu2021}. Among the elderly, an estimated $32\%$ experience gait disturbances, which significantly increase their risk of falls \cite{Mahlknecht2013}. In the United States and the United Kingdom, the annual cost of falls is estimated at $\$ 50$ billion \cite{Florence2018} and $\$ 2.6$ billion \cite{blogHumanCost}, respectively. Ankle exoskeletons offer a promising solution to counteract the decline in muscle strength and balance associated with aging, potentially improving gait and reducing the risk of falls \cite{Grimmer2019, Raitor2024}. Additionally, these devices may benefit a large group of patients suffering from conditions such as stroke, multiple sclerosis (MS), and cerebral palsy \cite{Shi2019, Androwis2021, Hunt2022}. Healthy individuals may still benefit from exoskeleton during high effort manual labour tasks \cite{Cho2018,Yan2021,Sado2019} and to reduce fatigue \cite{Wang2021, Wei2020}.\\

Exoskeletons can be broadly categorized as either active or passive, based on the presence of an external power source \cite{Tiboni2022}. Passive exoskeletons rely on mechanical components, such as springs, to assist walking by storing and releasing energy at specific points in the gait cycle \cite{Etenzi2020}. In contrast, active exoskeletons incorporate actuators like electric motors or pneumatic systems to directly reduce the physical effort required for walking, and in some cases, enable walking without muscle activity \cite{Slade2022, Chen2017}. For optimal support of user movements, active exoskeletons must accurately classify the user's motion intentions and reduce the latency between sensor input and control system response \cite{Tang2023_nano}.  To achieve this, a sensing system to record electrical activity, movements or forces must be designed.  Common sensors used in lower limb applications include IMUs, surface EMGs, and force or pressure sensors \cite{Wang2023}.  \\

The data from these sensors are processed, and motions  classified using machine learning techniques including Support Vector Machines (SVM), Convolutional Neural Networks (CNN), and Long Short Term Memory Networks (LSTM). Although SVMs have achieved near-perfect classification accuracy, CNNs have not yet reached comparable levels of accuracy \cite{Wang2023}. However, SVMs have limitations when it comes to training on large datasets or handling multi-class problems \cite{Cervantes2020}. A study by Li \textit{et al.} reported a classification accuracy of 95\% using CNNs and LSTMs, but their work was limited to isolated ankle movements rather than activities of daily living\cite{Li2023}. Cheng \textit{et al} obtained similar results with a comparable approach \cite{Cheng2020}. In contrast, Kim \textit{et al.} achieved an 88\% accuracy in classifying motions of daily living activities using CNNs trained on EMG and IMU data, which, while promising, is still insufficient for a seamless user experience \cite{Kim2023}. These studies highlight the potential of deep learning for motion classification, but do not achieve high enough accuracies  to enable seamless user experiences.\\

This study aims to assess whether scalable models, such as LSTM or CNN, can achieve high accuracy when used with a non-invasive recording system. We will evaluate both IMU and surface EMG signals  to determine their suitability for predicting ankle movements during activities of daily living. Electromyography (EMG) applications for long-term monitoring have traditionally been limited by the degradation of the skin-electrode interface, particularly when using conventional metal or gel electrodes. However, recent advancements in textile-based dry electrodes offer a promising alternative. These electrodes can be seamlessly integrated into clothing, providing enhanced comfort and durability for extended wear, while maintaining signal quality comparable to traditional methods \cite{Tang2023}. Additionally, to demonstrate model generalizability, we will test the models' performance in transfer learning scenarios. For user safety, we will also assess how the models perform in the event of individual sensor failure during operation. Furthermore, we provide a new publicly available dataset of EMG and IMU data, along with the data processing code and machine learning models, to support the development of new classification methods for exoskeleton control. \\

\section*{Results}
\subsection*{Data collection from human subjects}
\begin{figure}[ht]
\centering
\includegraphics[width=\linewidth]{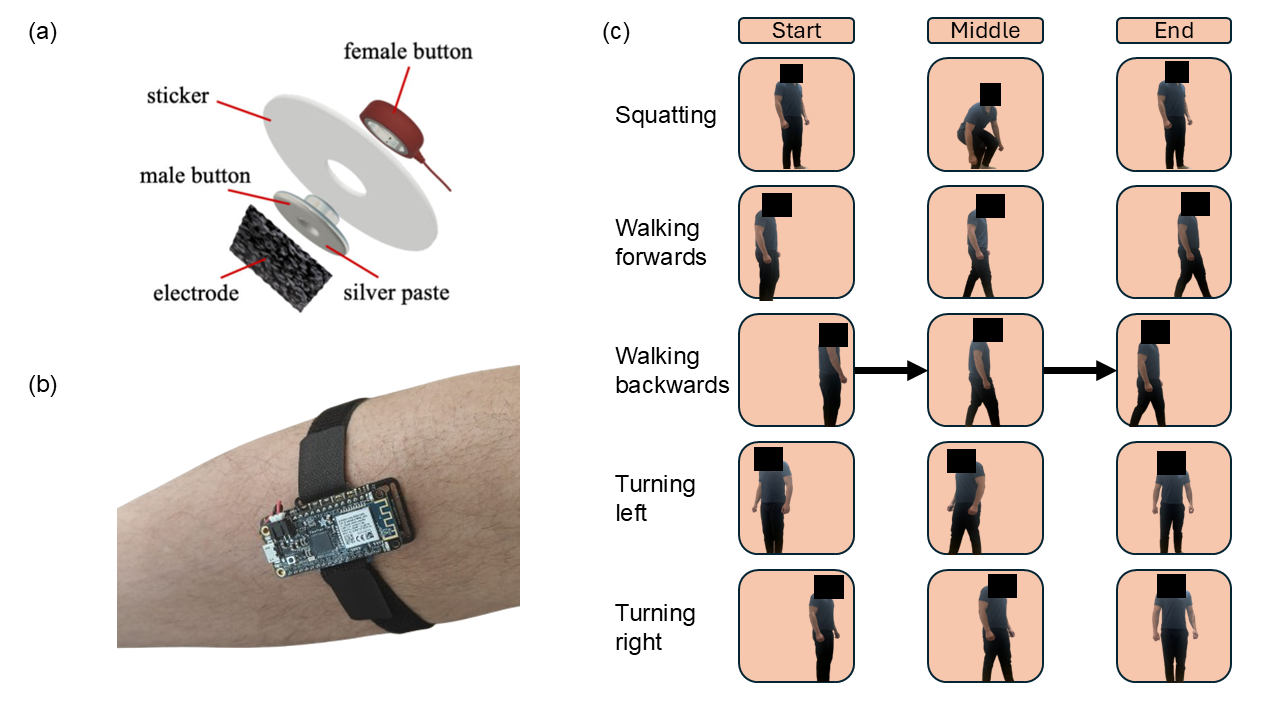}
\caption{\textbf{Data collection} (a) The schematic illustrates the custom-made electrode assembly used for EMG recordings \cite{Tang2023}. (b) Motion data was captured using commercial IMUs. (c) The dataset includes recordings from various activities: squatting, walking forwards and backwards, and turning left and right.}
\label{fig:setup}
\end{figure}

Ethical approval was obtained to record a total of 1,504 trials across three subjects using both EMG and IMU sensors (see Figure \ref{fig:setup}). IMU sensors were positioned on both shanks and on the right foot, while surface EMG sensors were placed to capture activity of the tibialis anterior, gastrocnemius medial and lateral heads, and soleus muscles. The five recorded motions, essential for navigating a barrier-free environment, included walking forwards, walking backwards, turning left, turning right, and squatting (to pick up an object or sit down). The collected signals were pre-processed and subsequently used to train a classifier for motion recognition.\\

\begin{figure}[ht]
\centering
\includegraphics[width=\linewidth]{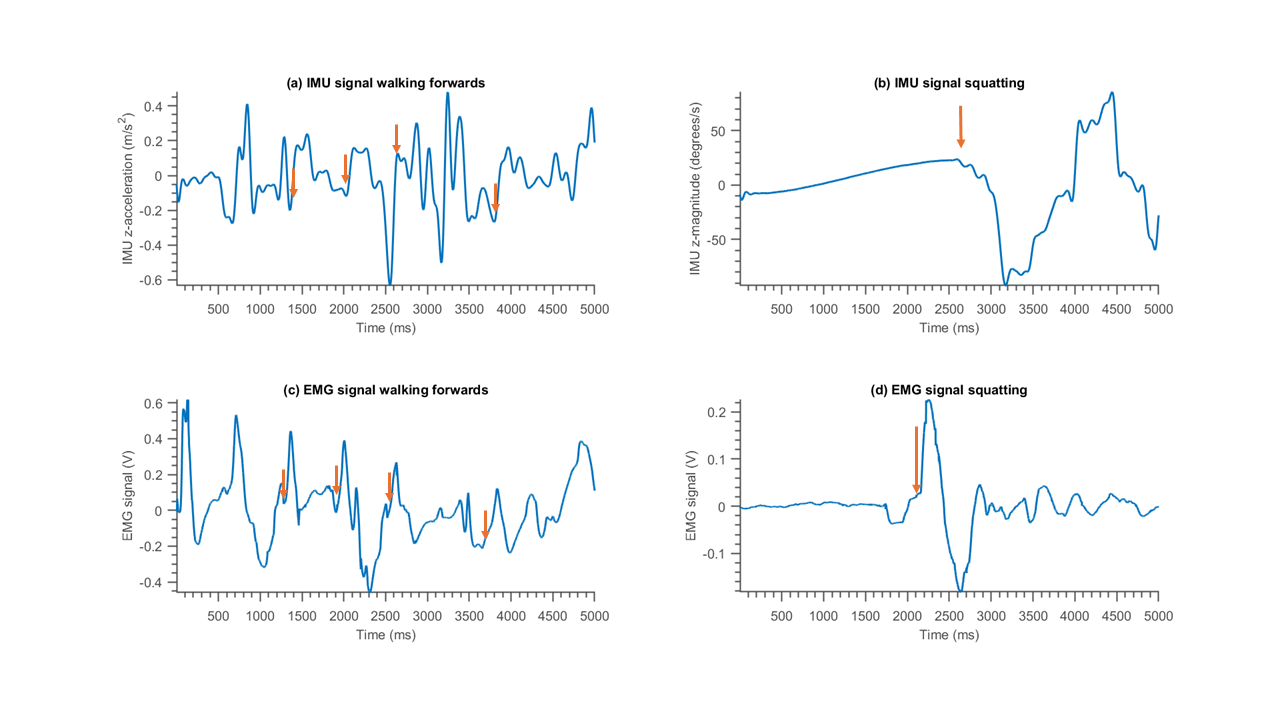}
\caption{\textbf{Signal recordings} Representative examples of the relationship between recorded signals and physiological processes (a-d). During forward walking, intermittent forward acceleration is detected by the IMU (a), which corresponds to regular activation of the soleus muscle (c). Similarly, during a squat movement, slight forward motion of the shank is observed by the IMUs, which is linked to activity in the tibialis anterior muscle (d). Orange arrows highlight key instances where the two signals align, emphasizing representative matches. The EMG signal leads the IMU data, as there is an inherent latency between the electrical signals indicating muscle contraction and the subsequent initiation of motion detected by the IMU.}
\label{fig:example_recordings}
\end{figure}

Representative raw recordings from two different trials, illustrating expected behavior, are shown in Figure \ref{fig:example_recordings}. During forward walking, the IMU’s vertical acceleration exhibits several distinct peaks, corresponding to most steps. However, some steps lack clear peaks, likely due to the subject not lifting their feet significantly during those strides. Concurrently, the soleus muscle recordings show multiple activations, as expected, given its role in plantar flexion during forward propulsion \cite{Lai2015}. In the squatting motion, activation of the tibialis anterior muscle is observed, which is essential for dorsiflexion to move the knees forwards \cite{Kim2017}. This activation is followed by a noticeable increase in forward horizontal acceleration, capturing the movement pattern.\\

\subsection*{Motion classification using deep learning}
\begin{figure}[ht]
\centering
\includegraphics[width=\linewidth]{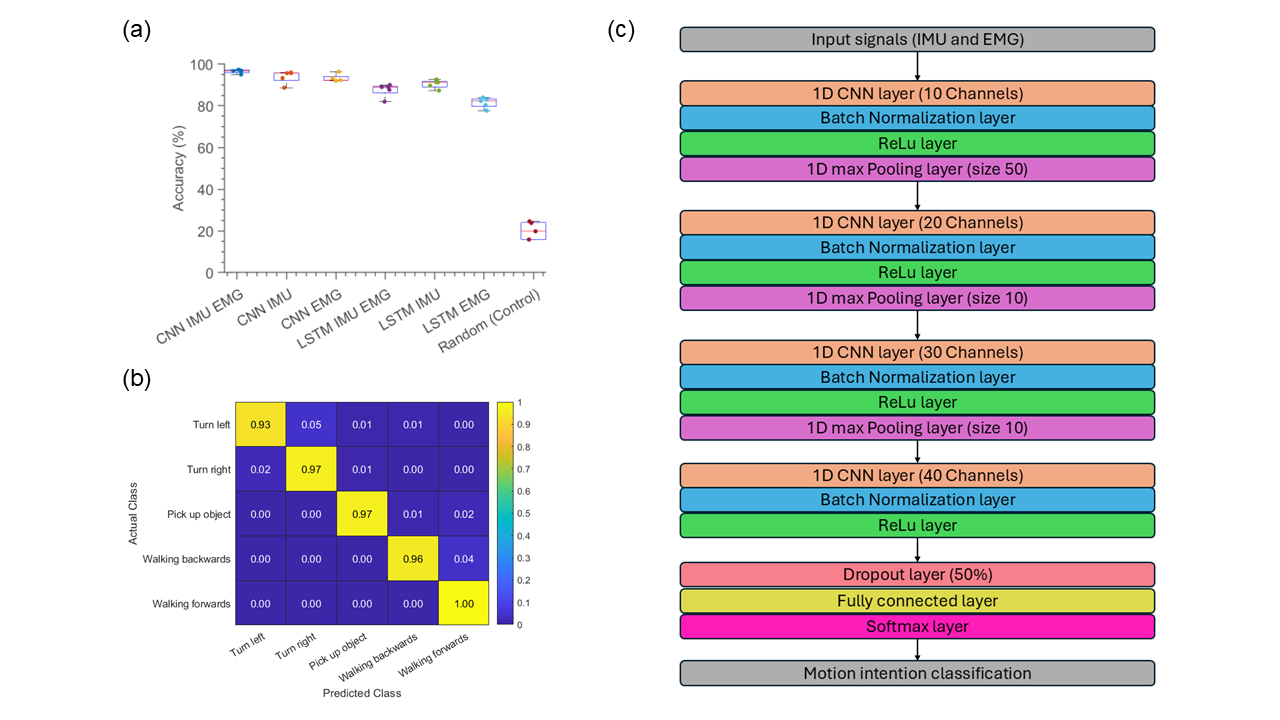}
\caption{\textbf{Classifier performance} (a) Performance of LSTM and CNN trained on different datasets (IMU only, EMG only and combined EMG and IMU). As a control, the accuracy of randomly choosing a class for each test sample was also reported. (b) Confusion matrix for the CNN trained on both IMU and EMG data. (c) Architecture of the CNN used for motion classification.}
\label{fig:cnn_perf}
\end{figure}

For classification, both LSTM and CNN models were trained on IMU data, EMG data and the combined IMU and EMG data. Combining IMU and EMG data for both the CNN and LSTM models results in high performance (see Fig. \ref{fig:cnn_perf}). The CNN demonstrated a slight performance advantage over the LSTM, achieving an accuracy of $96.5 \pm 0.8\%$ on the test data, compared to the LSTM’s $87.5 \pm 2.9\%$. As expected, models trained on individual modalities, such as EMG or IMU data alone, showed lower but still acceptable classification accuracies. Specifically, CNNs trained on only EMG data reached $93.9 \pm 1.6\%$, while those trained on only IMU data achieved $93.3 \pm 1.6\%$. Additionally, given that some users may require exoskeleton support for only one leg, we evaluated the classification accuracy using EMG and IMU data from a single leg. In this case, the CNN achieved an accuracy of $92.9 \pm 1.6\%$, highlighting the robustness of the model even with reduced input channels.\\

A detailed performance analysis of the CNN on test data using all channels is shown using a confusion matrix (see Fig. \ref{fig:cnn_perf}) and precision, recall and F1 scores for the five classes (see Table \ref{tab:scores}). The CNN demonstrates high accuracy across all classes, with only minor confusion observed between certain motions. Specifically, slight misclassification occurs between Turn Left and Turn Right, as well as between Walking Forwards and Walking Backwards. This confusion is likely attributable to the engagement of similar muscle groups and the symmetrical nature of leg movements, despite the differences in direction.\\

\begin{table}[ht]
\centering
\begin{tabular}{|l|l|l|l|}
\hline
 Motion & Precision & Recall & F1 \\
\hline
Turn left & 0.9757 & 0.9331 & 0.9539 \\
\hline
Turn right & 0.9483 & 0.9670 & 0.9576 \\
\hline
Pick up object & 0.9861 & 0.9658 & 0.9758 \\
\hline
Backwards & 0.9723 & 0.9565 & 0.9643 \\
\hline
Forwards & 0.9398 & 0.9969 & 0.9675 \\
\hline
\end{tabular}
\caption{\label{tab:scores}\textbf{CNN performance metrics.}}
\end{table}

\begin{figure}[ht]
\centering
\includegraphics[width=\linewidth]{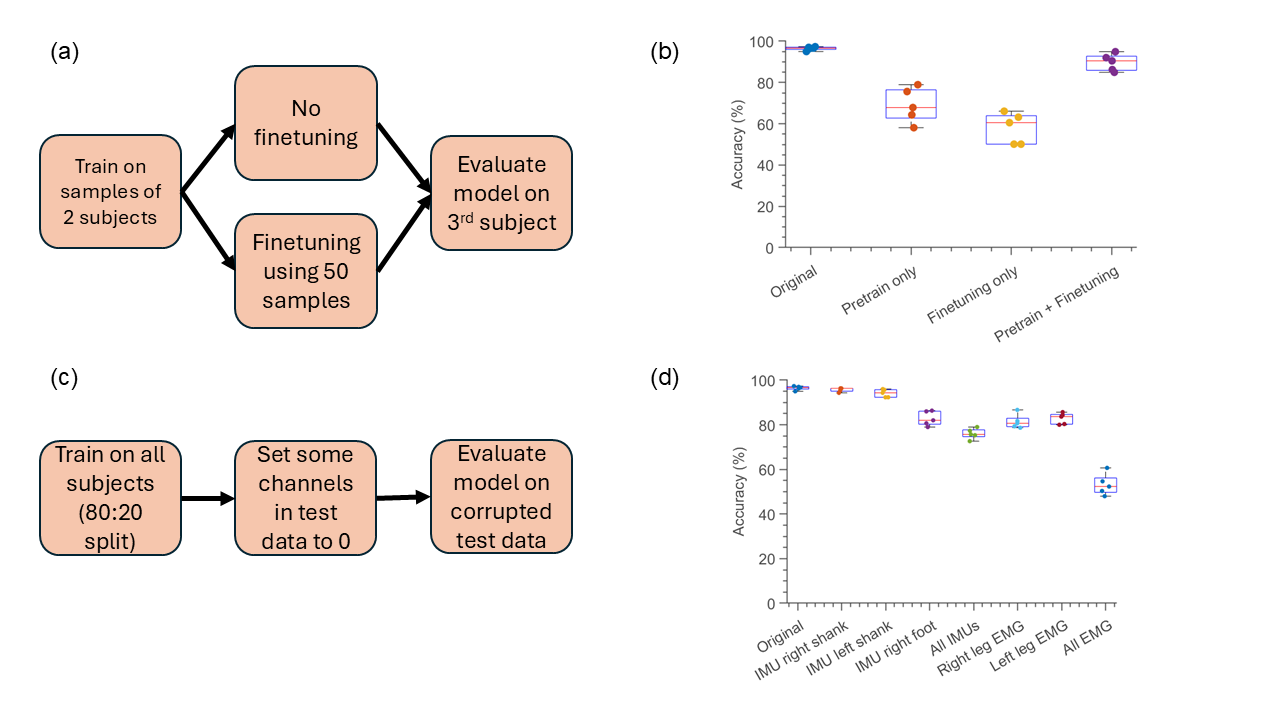}
\caption{\textbf{Transfer learning and model robustness.} (a) For transfer leaning the model is trained using data from 2 subjects only. The resulting model is then optionally fine-tuned using 50 samples before being evaluated on samples from a new subject. (b) Performance of the different models. (c) For model robustness the model is trained as previously but test data channels are corrupted. (d) The effect of different sensor corruptions on the classification results.}
\label{fig:transfer}
\end{figure}

\begin{figure}[ht]
\centering
\includegraphics[width=\linewidth]{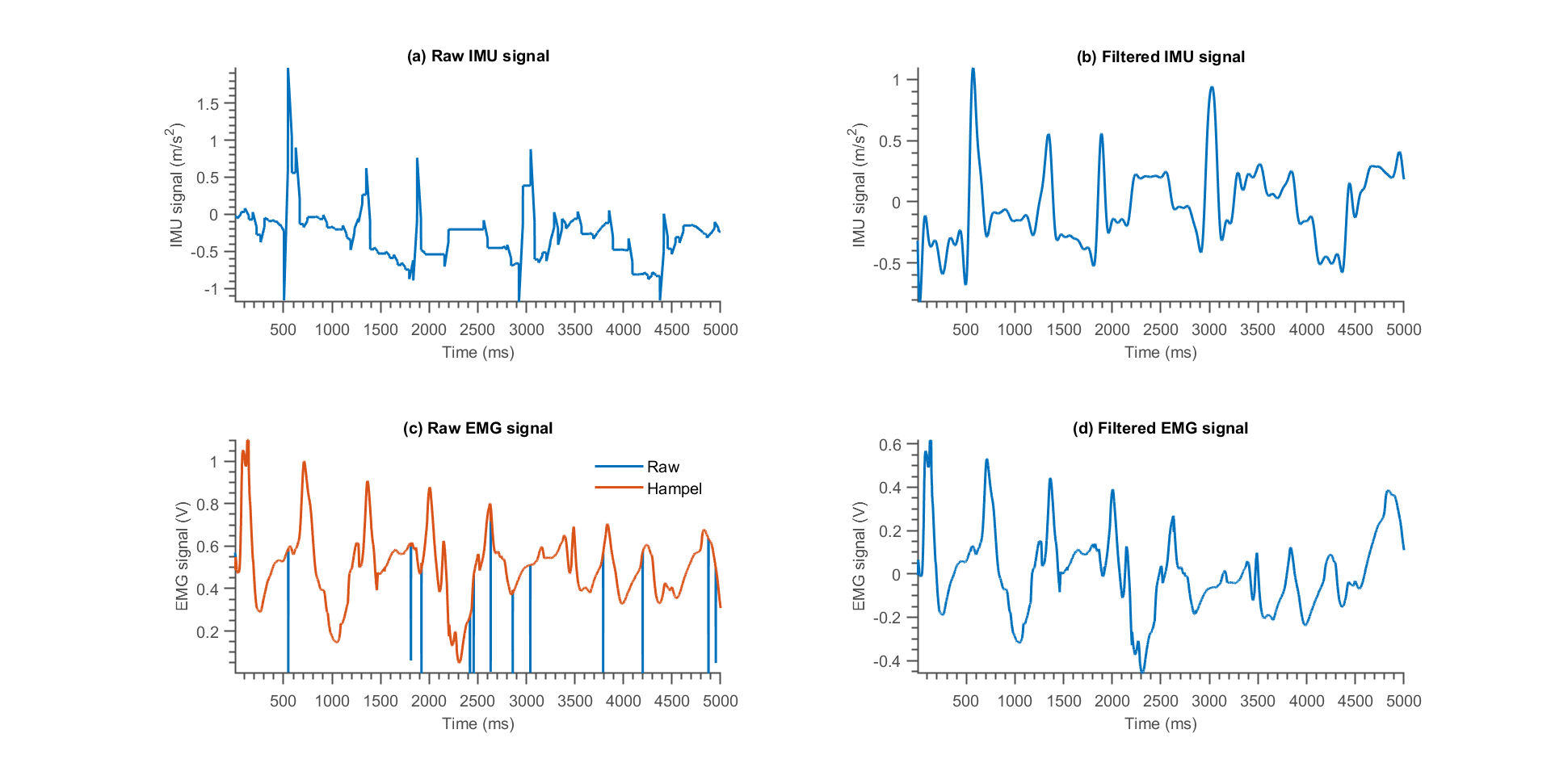}
\caption{\textbf{Signal processing.} Representative examples of EMG and IMU signal processing recorded during a forward walking trial. (a) The raw IMU signals are bandpass filtered to reduce noise (b). (c) EMG signals undergo outlier removal using a Hampel filter before bandpass filtering (d).}
\label{fig:sigproc}
\end{figure}

\subsection*{Transfer learning for model deployment}
 To demonstrate our model's ability to generalize to new users, we trained the CNN on data from only two subjects and evaluated its performance on the third subject.  As expected, the model's accuracy was notably lower $(69.0 \pm 7.6\%)$ (pretraining only) compared to the version trained on data from all three subjects $(96.5 \pm 0.8\%)$ (original) (see Fig. \ref{fig:transfer}). Additionally, training the model with only 10 samples per category from a new subject (finetuning only) resulted in very low accuracy of $58.0 \pm 6.7\%$. However, fine-tuning a pre-trained model using just 10 motion samples from the third subject achieved a performance of $89.7 \pm 3.7\%$.  For this, the convolutional layers were frozen after the pre-training, while the fully connected layer could still be modified. Part of the small difference is likely due to the original model being trained on 1,204 samples, compared to 1,051 for the transfer learning model. \\

\subsection*{Model robustness to sensor failure during operation}
During use, individual sensors may fail to provide accurate signals. For example, EMG electrodes might lose contact with the skin, or an IMU's battery could deplete. In such cases, the model's performance is likely to decrease due to the unexpected input from a faulty channel. However, it is crucial that the model maintains a sufficiently high level of performance to ensure the user's safety, allowing them to notify caretakers and retain some mobility until the sensors can be replaced. To assess the model's safety under these conditions, we evaluated classification accuracy on test samples where all channels of a particular sensor were set to zero, simulating absent signals, while the training samples remained unaltered. The model maintained an accuracy above $80\%$ even when individual IMU sensors failed or when EMG signals from one leg were lost (see Fig. \ref{fig:transfer}). The most significant performance drop occurred when the IMU on the foot was deactivated, likely due to the limited availability of redundant signals. In this scenario, the classification accuracy fell to $82.8 \pm 2.9\%$, which is notably lower than the accuracies observed when the IMUs on the left and right shanks were deactivated, yielding $94.1 \pm 1.6\%$ and $95.7 \pm 0.8\%$, respectively. This suggests that the foot IMU plays a critical role in providing unique data for accurate classification.\\

\section*{Discussion}
We demonstrate that EMG and IMU signals can be effectively utilized for high-accuracy motion intention prediction using a CNN. The proposed prototype is cost-efficient, built with commercially available IMU sensors and EMG electrodes, with minor modifications to the contact pads. Additionally, the machine learning models are lightweight, eliminating the need for external graphics cards during operation. Our model slightly outperforms comparable models for lower limb motion classification reported in the literature (see Table \ref{tab:literature_models}). \\

\begin{table}[ht]
\centering
\begin{tabular}{|l|l|l|l|l|l|}
\hline
 Study & Classes & Movements & Sensors& Model&  Accuracy\\
 
\hline

Kim \textit{et al.}   & 6 & Full body & IMU + EMG & CNN& 88.0\%\\
\hline
Li \textit{et al.} & 6 & Ankle only & EMG &CNN-LSTM& 95.7\%\\
\hline
Si \textit{et al.}  & 5 & Full body & EMG &CNN& 95.5\% \\
\hline
This work  & 5 & Full body & IMU &CNN& 93.3\%\\
\hline
This work & 5 & Full body & EMG &CNN& 93.9\%\\
\hline
\bf{This work}  & 5 & Full body & IMU + EMG &CNN& \textbf{96.5}\%\\
\hline

\end{tabular}
\caption{\label{tab:literature_models}Performance of similar machine learning models in the literature.}
\end{table}

Importantly, unlike the work by Li \textit{et al.}, our study involved full-body movements rather than limiting classification to isolated ankle motions, which are typically required for exoskeleton support \cite{Li2023}. This distinction makes direct comparisons challenging, as the movement classes we trained for are substantially different. The most comparable model in terms of sensor setup is that of Kim \textit{et al.}, which also focused on motions needed for navigating an environment but achieved only 88\% accuracy \cite{Kim2023}. Additionally, the model by Si \textit{et al.} achieved accuracy comparable to ours using only EMG recordings, but it focused on movements not commonly performed in daily life, such as a straight leg lift, which involve very different muscle activation patterns which likely facilitate classification \cite{Si2022}. \\

The accuracy of our CNN trained solely on the EMG portion of the dataset is comparable to results reported in other studies (see Table \ref{tab:literature_models}). Notably, the model also demonstrated strong performance when trained using only the IMU data. This is particularly promising, as muscle atrophy in paralyzed patients can lead to altered or diminished EMG signals, making IMU data a valuable alternative \cite{Huang2021}. These findings underscore that while each modality—EMG or IMU—can be effective for motion classification on its own, sensor fusion should be prioritized for more accurate and robust system development.\\

To the best of our knowledge, this is the first approach to incorporate textile electrodes in wearable exoskeleton control systems, which offer greater promise for long-term monitoring due to their comfort and wearability. Unlike previous studies, we specifically focused on evaluating model performance for motions that are most critical to exoskeleton operation. Furthermore, we conducted additional analyses to assess the feasibility of deploying our model in real-world settings, as well as its safety for users, ensuring that the system is both practical and secure for continuous use in assistive devices.\\ 

To effectively deploy an exoskeleton to new users, it is essential that the model generalizes well beyond the initial training dataset. Retraining the model for each individual would be both time-consuming and costly. However, previous studies have shown that fine-tuning with a small number of samples can still yield high-accuracy gesture classification \cite{Tang2023, CoteAllard2019}. In our study, we trained the model on a minimal dataset of just two subjects and then used fine-tuning to generalize the model to a third subject. With minimal calibration, the system achieved high accuracy, suggesting that transfer learning is a viable approach for our model, enabling effective adaptation to new users with minimal data and effort.\\

Our model robustness testing indicates a high level of user safety in the event of unexpected sensor failure during operation. We evaluated extreme scenarios, including the complete failure of one or more sensors. Although model performance drops noticeably under these conditions, the accuracy remains sufficient for the system to capture the correct motion in real-time, albeit with some minor difficulties—such as needing to repeat movements until the model makes the correct inference. The performance drop could likely be mitigated further by augmenting the training data with examples of faulty channels. However, since it is impractical to anticipate every possible mode of sensor failure, we opted to use a zero-signal approach as a proof-of-concept example of an unencountered faulty signal. This demonstrates the model's ability to handle sensor anomalies it was not explicitly trained on.\\

In conclusion, we present a robust system for accurate motion classification using surface EMG and IMU signals, with the highest accuracy achieved by a CNN that integrates both data modalities. The model shows strong generalizability to new users, requiring only minimal fine-tuning to maintain high performance. It also demonstrates resilience to individual sensor failures, continuing to deliver reliable accuracy even under suboptimal conditions. Furthermore, we validate the use of towel-based electrodes for motion intention applications, reinforcing the findings of Tang \textit{et al.}, who demonstrated accurate digit classification of quasistatic movements such as hand gesture recognition using this innovative electrode system \cite{Tang2023}. To support further research and development, we have made our code for data collection and analysis, as well as our datasets, publicly available—enabling the development of similar systems and machine learning models across the field.\\

The study has several limitations. First, we evaluated only a limited set of movements, which are adequate for navigating barrier-free environments, but further testing should encompass more complex motions, such as stair ascent and descent. For comprehensive exoskeleton control, it will also be necessary to predict joint angles in addition to classifying movements, which will require the collection of additional ground-truth data. LSTM-CNN hybrid models have shown promising results in predicting joint angles and could be explored for this purpose \cite{Zhu2022}. Another limitation is that the system was tested on only three subjects, although the transfer learning results indicate good generalizability. Moreover, a fourth IMU on the left foot should ideally be added for a more comprehensive dataset. Additionally, the current setup requires the attachment of multiple EMG sensors, which can be time-consuming. This challenge could be addressed by integrating the textile electrodes into a comfortable, wearable leg sleeve, streamlining the setup process. Future research will focus on collecting larger, more diverse datasets, including joint angle measurements, and integrating the system into an exoskeleton to provide enhanced support for users in real-world environments.
\\
\section*{Methods}
\subsection*{EMG sensor fabrication}
The conventional wet gel electrodes from a commercial three-electrode gel EMG setup (SeeedStudio EMG Detector) were replaced with custom-made textile electrodes, utilizing a graphene/PEDOT:PSS composite material \cite{Tang2023}. The electrodes were fabricated using commercial cotton towels (NU249W, Aston Pharma), which were first cleaned by treatment with ethanol and UV/ozone. The cleaned towels were then soaked in a graphene solution for 30 minutes, followed by drying at 120°C for an additional 30 minutes. This process was repeated with an aqueous solution of PEDOT:PSS (ph1000, Ossila, diluted 1:10). The alternating cycles of soaking in graphene and PEDOT:PSS were repeated several times to create the final composite material.

\subsection*{Data collection}
The EMG detector boards were connected to an ESP32-S3 microcontroller development board, which transmitted the data wirelessly over WiFi to a laptop. To measure motion, open-source commercially-available EmotiBit sensor units including 9-axis IMUs were employed \cite{emotibit}. The firmware for the IMU microcontrollers was manually developed using the manufacturer's software \cite{githubReleasesEmotiBitofxEmotiBit} and the Espressif 32 functionality provided by PlatformIO. Data was wirelessly transmitted using the EmotiBit Oscilloscope Software, which streamed the data locally over a UDP stream. The dataset comprises five distinct classes of motion (Turn left (90°), Turn right (90°), Picking up an object from the floor, Walking forwards and Walking backwards), with each class containing approximately 100 samples per subject, ensuring equal distribution across the dataset. 
\subsection*{Human subject study}
 Ethical approval for the study was granted by the University of Cambridge Department of Engineering Research Ethics Committee (registration number \#394). All participants provided informed consent prior to the experiment. Data collection took place in the Human Performance Laboratory at the University of Cambridge. The study involved three young healthy volunteer participants (2 females, 1 male), with an average age of $23 \pm 1$ years, height $173 \pm 9$ cm, and weight $69 \pm 16$ kg.
 EMG data were collected bilaterally for tibialis anterior, gastrocnemius medial and lateral heads, and soleus muscles. The electrodes were placed according to the SENIAM guidelines \cite{seniamRecommendationsSensor}. The 3 IMU units were placed on the right and left shanks, and right foot.

\subsection*{Timestamp matching between modalities}
The raw UDP stream from the EmotiBit oscilloscope was divided into three distinct files, each corresponding to data from a specific sensor. These files were processed using the EmotiBit DataParser, producing output files for the nine modalities associated with each IMU. The timestamps of the two recorded signals were then aligned to ensure synchronization. To match the frequency of the EMG signal, the IMU data were upsampled by a factor of 40 using linear interpolation. The final output for each trial was a CSV file, containing 35 columns and 5000 rows, representing a 5-second recording sampled at 1000 Hz.\\

\subsection*{Signal pre-processing}
Signal processing was performed in MATLAB (version R2023b).
The raw EMG values were converted to voltage $V$ using the equation
\begin{equation}
    V=\frac{1.1}{4095} \cdot S
\end{equation}
where $S$ are the raw signals and $V$ is the output voltage in Volt, knowing that the on-board analog-to-digital converter (ADC) featured a 12-bit resolution and a reference voltage of 1.1 V. The signals were then passed through a Hampel filter that removed outliers deviating by more than three standard deviations from the 50 surrounding samples (see Fig. \ref{fig:sigproc}) \cite{Bhowmik2017}. The signals were then bandpass filtered using a fifth order Butterworth filter with cutoff frequencies of 0.2 and 400 Hz \cite{DeLuca2010}. The signals were then normalised to zero mean and unit variance. Similarly, the raw IMU signals were bandpass filtered using a fifth-order Butterworth filter, but with cutoff frequencies of 0.2 Hz and 10 Hz, chosen based on the anticipated duration of the movements performed during the trial. These IMU signals were then normalized following the same procedure as the EMG signals.

\subsection*{Machine learning models}
All models were trained using the MATLAB Deep Learning Toolbox (version 23.2). The dataset, consisting of recordings and labels, was randomly split into training $(80\%)$ and testing $(20\%)$ sets. The CNN architecture was adapted from a motion classification model by Tang \textit{et al.} \cite{Tang2023}, and comprises several stages of 1D convolutional layers, each followed by batch normalization, ReLU activation, and max-pooling layers. These are followed by dropout layers, fully connected layers, and a softmax layer for classification. A detailed schematic of the CNN architecture is shown in Figure \ref{fig:cnn_perf}, while the model parameters are outlined in Table \ref{tab:parameters}. The LSTM model was built using two LSTM layers, each containing 100 units, followed by a $20\%$ dropout layer, a fully connected layer, and a softmax layer. This network structure is based on the two-layer LSTM example provided in the MATLAB documentation \cite{MatlabNet}.\\

Both models were trained using cross-entropy loss and optimized with the ADAM optimizer, using a batch size of 50 over 15 epochs. Performance metrics included accuracy, precision, recall, and F1 score. Training was conducted on the CPU of a Dell XPS 13 laptop (Intel i7), with individual CNN training times not exceeding 5 minutes. Each model was trained using five different random seeds, and performance metrics were averaged across these runs.\\

\begin{table}[ht]
\centering
\begin{tabular}{|l|l|l|}
\hline
 Layer (type)& Output shape & Parameter number\\
\hline
Cov-1D-1& [50, 10, 5000] & 3160\\
\hline
BatchNorm  & [50, 10, 5000] & 20\\
\hline
MaxPool1D & [50, 10, 100] & 0\\
\hline
Cov-1D-2& [50, 20, 100] &  1820 \\
\hline
BatchNorm& [50, 20, 100] &  40\\
\hline
MaxPool1D &[50, 20, 10] & 0\\
\hline
Cov-1D-3& [50, 30, 10] & 5430 \\
\hline
BatchNorm & [50, 30, 10] & 60\\
\hline
MaxPool1D &[50, 30, 1] & 0\\
\hline
Cov-1D-1& [50, 40, 1] & 10840\\
\hline
BatchNorm & [50, 40,1] & 80\\
\hline
Dropout & [50, 40, 1] &0\\
\hline
FullyConnected & [50, 5] & 205\\
\hline
\textbf{Total parameters} &- & \textbf{21655}\\
\hline
\textbf{Trainable parameters} & - & \textbf{21655}\\
\hline
\textbf{Non-trainable parameters} & - & \textbf{0}\\
\hline

\end{tabular}
\caption{\label{tab:parameters}Calculation of parameters in the CNN. The output shape is given as [batch size, number of channels, channel length].}
\end{table}
\subsection*{Statistical analysis}
Model performance was evaluated using accuracy, precision, recall, and F1 scores, as outlined in Hicks \textit{et al.} \cite{Hicks2022}. Accuracy, the most straightforward metric, represents the fraction of correctly classified samples and is calculated as:

\begin{equation} \text{Accuracy} = \frac{TC}{TC + FC} \end{equation}

where $TC$ is the number of true classifications, and $FC$ represents false classifications across all samples. This metric is particularly suitable for our evaluation because our dataset is balanced, ensuring that accuracy provides a meaningful overall assessment.

Precision measures how many of the samples predicted to belong to a particular class actually do, and is given by:

\begin{equation} \text{Precision} = \frac{TP}{TP + FP} \end{equation}

Here, $TP$ represents true positives, and $FP$ denotes false positives for a specific class. Precision is useful in cases where minimizing false positives is critical.

Recall, on the other hand, quantifies the proportion of actual class members that were correctly identified, and is defined as:

\begin{equation} \text{Recall} = \frac{TP}{TP + FN} \end{equation}

In this case, $FN$ refers to false negatives, or instances where the model failed to identify the true class. Recall is valuable when minimizing false negatives is a priority.

Finally, the F1 score combines precision and recall into a single metric that balances both, particularly useful when there is an uneven trade-off between the two. The F1 score is calculated as:

\begin{equation} \text{F1} = \frac{2 \cdot \text{Precision} \cdot \text{Recall}}{\text{Precision} + \text{Recall}} \end{equation}

This harmonic mean provides a robust measure when dealing with imbalanced class distributions or when both precision and recall are equally important in evaluating model performance.

\bibliography{references}

\subsection*{Code availability}
The underlying code for this study is available on GitHub and can be accessed via this link \url{https://github.com/Sr933/Exoskeleton-Data-Acquisition-and-Processing-Code}.

\section*{Data availability}
The datasets generated and/or analysed during the current study are available in the Apollo repository and can be accessed via this link \url{https://doi.org/10.17863/CAM.113504}.

\section*{Acknowledgements}
This study was partially funded by The MathWorks, Inc. The funder played no role in study design, data collection, analysis and interpretation of data, or the writing of this manuscript. 

\section*{Author contributions statement}
L.O. and J.M came up with the idea for the project. S.R.E, J.M. and D.K. designed the sensor system and collected the datasets. S.R.E. and C.T. performed the data analysis and designed the ML models. S.R.E drafted the first version of the manuscript.  L.O. provided the overall guidance and resources for the project. All authors reviewed the manuscript.

\section*{Competing interests}
The authors declare no competing interests.

\end{document}